\documentclass[aps,prl,reprint,groupedaddress]{revtex4-2}

\usepackage{graphicx}%
\usepackage{multirow}%
\usepackage{amsmath,amssymb,amsfonts}%
\usepackage{amsthm}%
\usepackage{mathrsfs}%

\usepackage{hyperref}
\usepackage{scholax}
\usepackage[scaled=1.075,ncf,vvarbb]{newtxmath}
\usepackage{mathrsfs}
\usepackage{siunitx}
\usepackage{gensymb}
\usepackage{setspace}

\newcommand{\mrm}{\mathrm}

\newcommand{\ket}[1]{{\left| {#1} \right\rangle}}
\newcommand{\bra}[1]{{\left\langle {#1} \right|}}

\newcommand{\avg}[1]{\left\langle {#1} \right\rangle }

\newcommand{\Esca}{\mathcal{E}}

\newcommand{\Bsca}{\mathcal{B}}

\begin{document}

    \title{Wideband Search for Axionlike Dark Matter Using Octupolar Nuclei in a Crystal}

\author{Mingyu Fan}
\author{Bassam Nima}
\author{Aleksandar Radak}
\author{Gonzalo Alonso-\'Alvarez}\thanks{Present address: Instituto Galego de F\'{i}sica de Altas Enerx\'{i}as, Universidade de Santiago de Compostela, 15782 Santiago de Compostela, Galicia, Spain.}
\author{Amar Vutha}
\affiliation{Department of Physics, University of Toronto, Toronto, Ontario M5S 1A7, Canada}
\date{\today}

\begin{abstract}
Most of the matter in the Universe is in the form of dark matter, which has evaded detection so far. Ultralight axionlike particles (ALPs) are a class of dark matter candidates that produce measurable signatures in the form of oscillating violations of discrete symmetries in nuclei. We report results from a search for an oscillating parity-odd time-reversal-odd nuclear Schiff moment of $^{153}$Eu ions in a crystal, which leads to constraints on ALP-gluon coupling strength across a wide band spanning eight decades in ALP mass.
\end{abstract}
\maketitle

Astrophysical observations indicate that the majority of matter in the Universe is dark~\cite{Collaboration2020b}. However, dark matter has only been observed to interact gravitationally with ordinary matter, and no other interactions with standard model particles have been measured in experiments~\cite{bertone2018new}. 

Ultralight axionlike particles (ALPs), originally proposed in Refs.\ \cite{Preskill:1982cy,Abbott:1982af,Dine:1982ah}, are a well-motivated and viable model of cold dark matter in galaxies~\cite{Hui2021,ferreira_ultra-light_2021,chadha-day_axion_2022}. When ALPs interact with the gluons in atomic nuclei, they induce oscillating nuclear moments that are odd under parity (P) and time-reversal (T) symmetries \cite{stadnik_axion_2014}. Experiments sensitive to P-odd T-odd nuclear moments -- such as searches for permanent electric dipole moments (EDMs) of particles -- can detect these oscillations. For instance, broadband laboratory bounds on the ALP-gluon coupling strength in the mass range below $3\times10^{-12}$ eV have been obtained from neutron EDM experiments~\cite{abel_search_2017, schulthess_new_2022} and the HfF$^+$ electron EDM experiment~\cite{roussy_experimental_2021}.  
In addition to laboratory experiments, gluon-coupled ALPs can also affect cosmological processes like big bang nucleosynthesis~\cite{Blum:2014vsa}, and be sourced by astrophysical bodies like neutron stars~\cite{Hook:2017psm} and white dwarfs~\cite{Balkin:2022qer}.
All the above search methods mainly test ALP models with enhanced gluon couplings compared to canonical QCD axion models~\cite{Hook:2018jle,DiLuzio:2021pxd}. 

In this Letter, we probe oscillatory P-odd T-odd nuclear moments using precision spectroscopy of octupolar nuclei in a crystal. Our measurement method builds upon the search for static nuclear T violation proposed in Ref.~\cite{ramachandran_nuclear_2023}. We use $^{153}$Eu (nuclear spin $I=5/2$), a stable isotope that has a collectively enhanced P-odd T-odd nuclear Schiff moment induced by ALPs~\cite{flambaum_enhanced_2020, sushkov_schiff_2024}. 

Eu$^{3+}$ ions doped into a yttrium orthosilicate crystal (Eu:YSO) are located at noncentrosymmetric crystal sites, where their charge distribution is strongly electrically polarized by the neighboring ions in the crystal. In these trapped, polarized europium ions, the Schiff moment of the $^{153}$Eu nucleus interacts with the electron density gradient, causing nuclear spin states to shift their energy. The characteristic feature of these energy shifts produced by the Schiff moment is their dependence on the relative orientation of the nuclear spin vector, $\vec{I}$, and the electric dipole moment of the polarized Eu$^{3+}$ ion, $\vec{D}$. Observation of an \emph{oscillating} energy shift proportional to $\hat{I} \cdot \hat{D}$ is the signature of an oscillatory P-odd T-odd nuclear moment, as would be produced by ALPs.

\begin{figure*}
    \centering
    \includegraphics[width=0.75\linewidth]{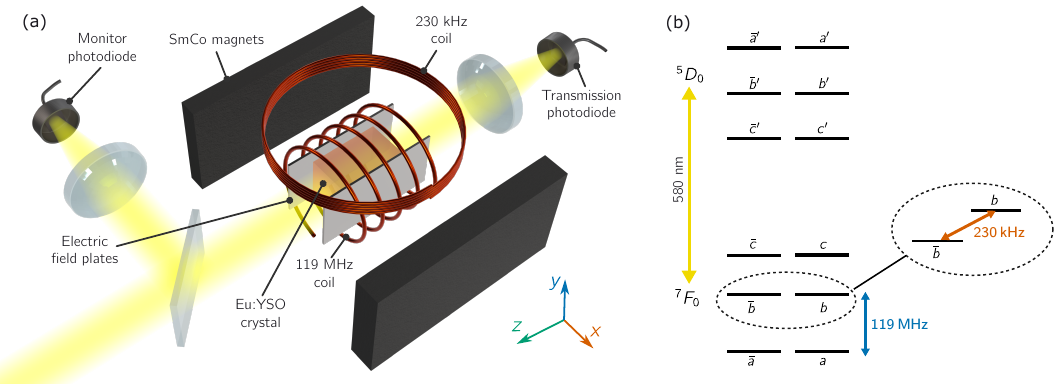}
    \caption{(a) Apparatus schematic. The ALP search experiment used laser absorption spectroscopy of the $^7F_0 \to {}^5D_0$ transition in an Eu:YSO crystal. 
    (b) Energy levels of $^{153}$Eu$^{3+}$:YSO. The ground ${}^7F_0$ and excited ${}^5D_0$ electronic states are connected by a $\SI{580}{\nano\meter}$ optical transition. Within each electronic state, the six nuclear spin sublevels are split into three pairs of Kramers doublets, with opposite nuclear spin orientations for the two sublevels in each doublet. The $b-\bar{b}$ nuclear spin transition at $\SI{230}{\kilo\hertz}$ was probed using precision rf spectroscopy.
    }
    \label{fig:setup}
\end{figure*}

Eu:YSO contains statistically equal numbers of Eu$^{3+}$ sites with oppositely directed $\vec{D}$ vectors, distributed throughout the crystal. These sets of oppositely polarized ions form two ensembles whose ALP-induced energy shifts oscillate exactly out of phase. Meanwhile, the shifts of these two ensembles due to magnetic fields (including stray fields and spurious backgrounds) are identical and in phase. Therefore, comparing measurements between oppositely polarized ions trapped within the same crystal provides a means to separate ALP signals from magnetic-field-induced systematic errors. Accurate cancellation of magnetic field effects using this method has been demonstrated in Ref.\ \cite{nima_precision_2025}. 

Furthermore, the narrow ${}^7F_0\to {}^5D_0$ transition of Eu$^{3+}$ enables optical control of the europium ions, which leads to fast state preparation, low-noise readout of the nuclear spin states, and simultaneous interrogation of the oppositely polarized ensembles. Finally, the large number of Eu$^{3+}$ ions available in the crystal results in high sensitivity to ALP signals.

To quantify the sensitivity, we note that the ALP dark matter field performs nearly coherent oscillations at a frequency $f_\mrm{ALP}=m_ac^2/h$, where $m_a$ is the ALP mass, with a small frequency spread due to the velocity dispersion of dark matter in the Galactic halo. The oscillating ALP field produces a time-varying value of the dimensionless $\theta$ parameter of QCD~\cite{stadnik_axion_2014},
\begin{equation}
\label{eq:theta}
    \theta(t) = \frac{C_G}{f_a} a(t).
\end{equation}
$\theta$ in this equation is the physical CP-violating parameter including the phase of the determinant of the quark mass matrix, $C_G$ is the gluon coupling constant ($C_G=1$ for the QCD axion)~\cite{roussy_experimental_2021}, $f_a$ is the ultrahigh energy scale associated with ALP formation, and $a(t)$ is the ALP field at the location of the experiment. The amplitude of the ALP field is estimated, assuming that ALPs are the sole source of dark matter, by setting the local dark matter (DM) energy density, $\rho_{\rm DM} = m_a^2 \avg{a^2}$. 

The ALP-induced $\theta$ parameter in turn creates a time-varying nuclear Schiff moment of $^{153}$Eu given by
\begin{equation}
    \vec{\mathscr{S}}(t) = \mathscr{S}_{\theta} \frac{\vec{I}}{|I|} \, \theta(t).
    \label{eq:schiff}
\end{equation}
To interpret our measurements of oscillatory Schiff moments in terms of $\theta(t)$, we use the value $\mathscr{S}_{\theta} \approx 0.15\, e \,{\mrm{fm^3}}$ that was estimated in Ref.\ \cite{sushkov_schiff_2024}. 

In Eu:YSO, the effective Hamiltonian describing the P-odd T-odd energy shifts of the nuclear spin states is $\mathcal{H}_\mrm{ALP} = \vec{\mathscr{S}} \cdot \vec{W}$, where $\vec{W}$ is a quantity proportional to the electron density gradient at the nucleus. The direction of $\vec{W}$ is parallel to that of the electric polarization of the Eu$^{3+}$ ion, $\hat{D}$. Calculations of the quantity $W$ for Eu$^{3+}$ ions in Eu:YSO indicate that $W\approx 10^{3}$ atomic units \cite{cheng_private_communication}. In the following discussion we use the reference value $W = 10^{3} \ \mrm{a.u.}$, where $1 \ \mrm{a.u.} = h \times 44.4$ Hz/e fm$^{3}$. 

Combining the above relations, the effective Hamiltonian for the nuclear spin degree of freedom of $^{153}$Eu$^{3+}$ ions in Eu:YSO is
\begin{equation}
\begin{split}
    \mathcal{H}_\mrm{ALP}(t) & = \hbar \, \Omega\, \hat{I}\cdot\hat{D}
\label{eq:T-violating-hamiltonian} \\
& = \left[ \left( h \times 6.7 \ \mrm{kHz} \right) \left(\frac{W}{10^{3} \ \mrm{a.u.}} \right) \, \theta(t) \right]  \hat{I}\cdot\hat{D} 
\end{split}
\end{equation}
In our experiment, we measure the time-dependent P-odd T-odd frequency shift due to $\mathcal{H}_\mrm{ALP}(t)$ to determine $\theta(t)$, and hence the ALP-gluon coupling strength, $C_G/f_a$.

A schematic diagram of the experiment is shown in Fig.~\ref{fig:setup}(a). The measurements reported here used a YSO crystal doped with $^{153}$Eu$^{3+}$ at 0.01\% concentration. The crystal, $3.5\times 4.0\times\SI{5.0}{\milli\meter}$, was attached to a cold plate and maintained at $\SI{5}{\kelvin}$ by a cryocooler. A pair of metal plates adjacent to the crystal were used to apply a static electric field, $\Esca_\mrm{dc}$, along the $\hat{x}$ axis (parallel to the dielectric $D_1$ axis of the crystal). A coil was used to apply radio-frequency (rf) magnetic fields along the $\hat{y}$ axis (parallel to the dielectric $D_2$ axis) at $\SI{230}{\kilo\hertz}$ to drive the $b \leftrightarrow \bar{b}$ transition for precision spectroscopy, shown in the energy level diagram of the Eu$^{3+}$:YSO system in Fig.~\ref{fig:setup}(b). Another coil produced rf magnetic fields along the $\hat{z}$ axis at $\SI{119.2}{\mega\hertz}$ to drive the $(a,\bar{a}) \leftrightarrow (b,\bar{b})$ nuclear spin transition in the $^7F_0$ electronic ground state for state preparation. SmCo permanent magnets were used to apply a bias magnetic field $\Bsca_\mrm{dc} \approx 350$~G across the $\hat{x}$ axis of the Eu:YSO crystal, in order to resolve the $a-b$ and $\bar{a}-\bar{b}$ resonances.
A $\hat{x}$-polarized laser propagating through the crystal along the $\hat{z}$ axis (parallel to the dielectric $b$ axis) was used for state preparation and measurement.

\begin{figure*}
    \centering
    \includegraphics{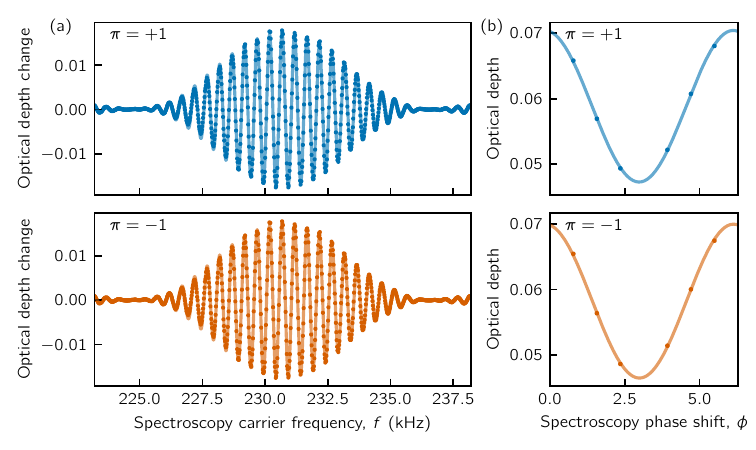}
    \caption{(a) Ramsey spectroscopy of the $b- \bar{b}$ transition for Eu$^{3+}$ ions of different polarizations $\pi=\pm1$. The vertical axis is the measured optical depth change between experiments with opposite spectroscopy pulse phase differences, $A(f,0^\circ) - A(f,180^\circ)$. Here the optical depth is defined as $\mrm{OD}=-\log(P_T/P_M)$, where $P_T$ ($P_M$) is the optical power measured at the transmission (monitor) photodiode. The solid lines show fits to the theoretical Ramsey line shape~\cite{ramsey_molecular_1985} with center frequency, pulse area, and amplitude being the only free parameters. (b) Phase scans of the $b- \bar{b}$ transition for different ion polarizations $\pi=\pm1$. The vertical axis is the optical depth measured with a fixed spectroscopy carrier frequency, $A(f=\SI{230.7}{\kilo\hertz},\phi)$. The solid lines show sinusoidal fits. The Ramsey phase signals for $\pi=\pm1$ are identical to within the uncertainty.}
    \label{fig:ramsey}
\end{figure*}

We probed ions at ``site 1'' in Eu:YSO~\cite{konz_temperature_2003}. The inhomogeneously broadened linewidth of the optical transition in our sample was $\SI{700}{\mega\hertz}$, and the linewidth of the spectral antiholes used for our measurements was typically 1 MHz. In the presence of the electric field $\Esca_\mrm{dc}$, the spectral antiholes for the oppositely polarized ion ensembles shift in opposite directions, determined by their respective orientations, $\pi\equiv \hat{D}\cdot\hat{x} = \pm1$. We used $\Esca_\mrm{dc} = \SI{80}{\volt\per\centi\meter}$, which produces a spectral shift $\Delta \nu = \mp \SI{2.2}{\mega\hertz}$~\cite{zhang_precision_2020} for the $\pi = \pm1$ ensembles. Thus the optical absorption features from oppositely polarized ions can be distinguished and $\Omega$ can be extracted. 

The experiment sequence begins with an optical pumping pulse that sweeps through the ${}^7F_0 \ a, \bar{a} \to {}^5D_0 \ c', \bar{c}'$ resonance~\cite{cruzeiro_characterization_2018} while  $\Esca_\mrm{dc}$ is on. This pulse clears out the population of spectroscopic classes that are unused in the experiment, and creates a flat background for absorption measurements. Next, $\Esca_\mrm{dc}$ is turned off, and optical pulses tuned to the ${}^7F_0 \ a, \bar{a} \to {}^5D_0 \ c', \bar{c}'$ and ${}^7F_0 \ c, \bar{c} \to {}^5D_0  \ b',\bar{b}'$ resonances pump ions into the ${}^7F_0 \ b, \bar{b}$ states using the spectral hole burning methods described in Refs.~ \cite{macfarlane_coherent_1987,cruzeiro_characterization_2018,nima_precision_2025}. 
Then a $\pi/2$ pulse is applied on the $b \leftrightarrow \bar{b}$ transition to equalize the populations of the two states, and remove any population imbalance from previous experiment cycles.  
An rf hyperbolic-square-hyperbolic pulse~\cite{tian_reconfiguration_2011} swept across the $b \leftrightarrow a$ transition moves ions out of the $b$ state, initializing ions in the $\bar{b}$ state. 

Two rf $\pi/2$ spectroscopy pulses at a carrier frequency $f$ and with relative phase difference $\phi$, separated by $T =\SI{2}{\milli\second}$, are applied to drive the $b \leftrightarrow \bar{b}$ transition. Following this, a second rf sweep over the $b \leftrightarrow a$ transition transfers ions from $b$ to $a$. All the above rf state-preparation and spectroscopy pulses act simultaneously on the $\pi=\pm1$ ensembles. We note that the lab electric field is switched off during the entire rf pulse sequence, eliminating electric-field-dependent systematic errors from the critical precision spectroscopy steps.

Finally, optical probe pulses tuned to ${}^7F_0 \ a, \bar{a} \to {}^5D_0 \ c', \bar{c}', \pi=\pm1$ resonances are used to measure laser absorption, yielding a signal $A(f,\phi)$ proportional to the population transferred to the $b$ state by the spectroscopy pulses. $\Esca_\mrm{dc}$ is switched on during this detection step, so that the optical absorption features from the $\pi = \pm1$ ions become separated by 4.4 MHz and can be distinguished~\cite{nima_precision_2025}. Fig.~\ref{fig:ramsey}(a) shows the signals measured with $\pi=\pm1$ ions as the spectroscopy carrier frequency $f$ is scanned across the $b - \bar{b}$ resonance.

To precisely measure the $b -\bar{b}$ resonance frequency, $f_0(b \bar{b})$, we use a modified Ramsey method \cite{Klein1987}. In each cycle, we measure $A(f,\phi)$ for six different values of $\phi$ between 0 and $2 \pi$ as shown in Fig.~\ref{fig:ramsey}(b). The corresponding absorption signal, $A(f,\phi) = A_0 \cos \left\{ \phi +2\pi \left[ f-f_0(b\bar{b}) \right] T \right\}$, is fit to a cosine to extract the phase offset $2\pi \left[ f-f_0(b\bar{b}) \right] T$, and from it the resonance frequency $f_0(b\bar{b})$. This procedure makes the measurement immune to systematic errors arising from asymmetries in the Ramsey line shape \cite{Klein1987}. The resonance frequencies measured for the $\pi=\pm1$ ensembles are denoted $f_0(b\bar{b}, \pi=\pm1)$.

We define the sum frequency $f_s$ and difference frequency $f_d$ for the two $\pi = \pm1$ ensembles as
\begin{align}
    f_s&=f_0(b\bar{b}, \pi=+1) + f_0(b\bar{b}, \pi=-1),\\
    f_d&=f_0(b\bar{b}, \pi=+1) - f_0(b\bar{b}, \pi=-1).
    \label{eq:f_d}
\end{align}
The difference frequency $f_d$ is 
\begin{equation}
    f_d = 4 \frac{\Omega}{2\pi} \, \bra{b}\hat{I}\cdot\hat{D}\ket{b},
\end{equation}
where $\Omega$ is the T-violation parameter defined in Eq.~\ref{eq:T-violating-hamiltonian}. We set $\hat{D} \parallel \hat{x}$ based on the the fact that the oscillator strength of the ${}^7F_0 - {}^5D_0$ optical transition is strongest for $\hat{x}$-polarized light \cite{cruzeiro_characterization_2018}. The nuclear spin projection of the $b$ state is calculated to be $\bra{b}\hat{I}\cdot\hat{D}\ket{b} =  0.25$, as discussed in Supplemental Material (SM) \cite{Supplemental}\nocite{longdell_hyperfine_2002,yano_nonlinear_1992,stone_table_2014, derevianko_detecting_2018, baxter_recommended_2021}.

The experiment acquired data between July 11-20, 2025, and a total of \SI{272808}{} pairs of $f_0(b\bar{b},\pi=\pm1)$ frequencies were measured. From these measurements, we consider the time-varying T-violating shift, $\delta f_d\equiv f_d - \mrm{avg} (f_d)$, that includes possible ALP signals. To search for ALPs in the frequency domain, we compute the $\delta f_d$ power spectrum using the generalized Lomb-Scargle periodogram~\cite{vanderplas_understanding_2018}, since the measurement times are not uniformly spaced. 

\begin{figure}[t]
    \centering
    \includegraphics{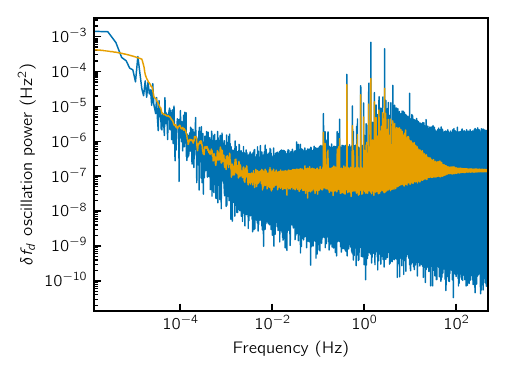}
     \caption{Best-fit oscillation power of the T-violation parameter $\delta f_d$ shown in blue. The peaks corresponds to technical noise frequencies and their aliased frequencies by experiment cycles. The ALP-free noise model is shown in yellow.}
    \label{fig:spectrum}
\end{figure}

The spectral power is calculated between $\SI{1.3}{\micro\hertz}$ (inverse of the experiment integration time, $T_\mrm{int} = 7.5 \times 10^5$ s) to $\SI{500}{\hertz}$ (inverse of the spectroscopy pulse time, $T_\mrm{Ramsey} = 2$ ms) with a frequency bin width of $1/ T_\mrm{int}$ (see SM) \cite{Supplemental}.
The power spectrum of $\delta f_d$ is shown in Fig.~\ref{fig:spectrum}. Noise peaks above the average noise of $1.6\times10^{-8}$ Hz$^2$ arise from three sources: (1) peaks at the cryocooler cycle frequency (\SI{1.40017}{\hertz}) and its harmonics, (2) slow variations in $\delta f_d$ below 1 mHz, and (3) technical noise due to a data acquisition device at 20.0000 Hz.

We use a statistical analysis similar to Ref.\ \cite{foster_revealing_2018} to determine the ALP-gluon coupling. We construct a $\delta f_d$ power spectrum model, $\mathcal{M}$, from the expected signal of an ALP with mass $m_a$ and ALP field amplitude $\theta_0$, as well as a non-ALP background noise model $\mathcal{B}$ (see SM for the details \cite{Supplemental}). The best-fit parameters for the background noise, $\hat{\mathcal{B}}$, are determined for each frequency $f_\mrm{ALP}$ from the average of the $\delta f_d$ power in a window of width $1\times10^{-4} f_\mrm{ALP}$ around $f_\mrm{ALP}$. This frequency range for power averaging is much broader than the ALP line shape, which has a width of $10^{-6} f_\mrm{ALP}$. If there is any ALP signal in the data, the noise model $\hat{\mathcal{B}}$ contains only 1\% of the signal power, and therefore it can be used as an ALP-free noise background for testing the presence of ALPs. The sum of the ALP signal and noise models is compared with the experimentally measured spectrum using the likelihood function
\begin{equation}
    \mathcal{L}(d|\mathcal{M}, \{m_a, \theta_0, \mathcal{B}\})=\prod_{k}\frac{e^{-S_k/\lambda_k(m_a, \theta_0, \mathcal{B})}}{\lambda_k(m_a, \theta_0, \mathcal{B})}.
    \label{eq:likelihood}
\end{equation}
Here $k$ iterates over power spectrum frequencies, $\lambda_k(m_a, \theta_0, \mathcal{B})$ is the model $\delta f_d$ oscillation power at frequency index $k$, and $S_k$ is the measured oscillation power at frequency index $k$~\cite{foster_revealing_2018}.

The maximum likelihood estimate of the dimensionless ALP field amplitude, $\hat{\theta}_0\equiv C_G \sqrt{2\avg{a^2}} /f_a$, is obtained using the above likelihood function. The likelihood function is also used to construct a 95\% upper limit, $\theta_0^{95\%}$, through the definition~\cite{foster_revealing_2018}
\begin{equation}
\begin{split}
    2 [&\log \mathcal{L}(d|\mathcal{M}, \{m_a, \theta_0^{95\%},\hat{\mathcal{B}}\}) \\  &- \log \mathcal{L}(d|\mathcal{M}, \{m_a, \hat{\theta}_0, \hat{\mathcal{B}}\})] = -\chi_\mrm{half}^2.
\end{split}
\end{equation}
Since the upper limit $\theta_0^{95\%}$ is always larger than the best fit value $\hat{\theta}_0$,  we use the 95\% critical value of a half-chi-squared distribution with one degree of freedom, $\chi_\mrm{half}^2=2.71$.

We perform this analysis for ALP oscillation frequencies from 1.3 $\mu$Hz to 500 Hz in $7.7\times10^7$ steps. The frequency step size is either smaller than the expected ALP signal linewidth, or smaller than the frequency resolution of the experiment when the ALP linewidth is unresolved. We compare the likelihood of the ALP signal with that of the noise model for each frequency, and find that all frequencies where the ALP signal model likelihood exceeds the global $3\sigma$ threshold match the known technical noise sources discussed above~\cite{Supplemental}. There is no statistically significant ALP signal in the dataset.

\begin{figure}[h!]
    \centering
    \includegraphics{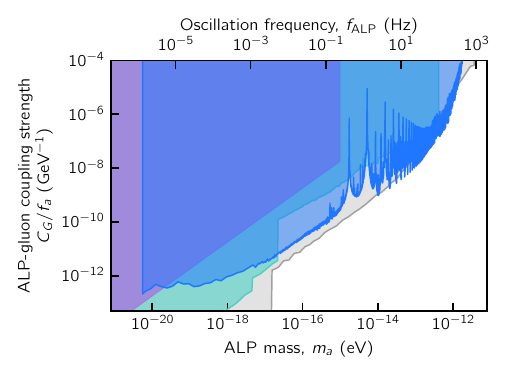}
    \caption{95\% confidence level exclusion range on the ALP-gluon coupling from this work shown as blue shaded regions. The previous experimental constraints in this range of ALP masses, compiled by Ref.~\cite{ohare_cajohareaxionlimits_2020}: trapped neutron and neutron beam EDM experiments~\cite{abel_search_2017, schulthess_new_2022} (gray), clock comparison experiments (green)~\cite{hees_searching_2016, kennedy_precision_2020, kobayashi_search_2022, sherrill_analysis_2023, zhang_search_2023, banerjee_oscillating_2023, kim_probing_2024, fuchs_searching_2025}, and the HfF$^+$ electron EDM experiment (purple)~\cite{roussy_experimental_2021}.}
    \label{fig:exclusion}
\end{figure}

The experiment integration time is shorter than the ALP coherence time for lighter ALPs that have $m_a\lesssim10^{-14}$ eV, so that we need to include a correction for stochastic fluctuations of the ALP field amplitudes~\cite{centers_stochastic_2021}: due to the heavy-tail Rayleigh distribution of the ALP field amplitude, the measured amplitude is statistically more likely to be lower than the true time-averaged amplitude, unless the experiment integration time is long enough to sample stochastic changes of the amplitude. We correct for the stochastic fluctuation effect using a Monte Carlo simulation (see SM). After applying the stochastic correction, we convert $\theta_0^{95\%}$ to a bound on $C_G/f_a$ using Eq.~\ref{eq:theta}, assuming the ALP field is the dominant source contributing to local dark matter with density $\rho_\mrm{DM} = 0.4$ GeV / cm$^3$~\cite{catena_novel_2010}. The resulting bound, with statistical variations smoothened using a rolling average, is shown in Fig.~\ref{fig:exclusion}. The final 95\% confidence upper bound on $\theta_0$ spans more than eight decades in ALP mass.

The wideband bound from Eu:YSO shown in Fig.\ \ref{fig:exclusion} sets the most stringent limit of any atomic~\cite{hees_searching_2016, kennedy_precision_2020, kobayashi_search_2022, sherrill_analysis_2023, zhang_search_2023, banerjee_oscillating_2023, kim_probing_2024, fuchs_searching_2025} or molecular~\cite{roussy_experimental_2021} experiment in this range of ALP masses. 
The published bound from neutron beam EDM experiments~\cite{schulthess_new_2022} is also shown for comparison in Fig.\ \ref{fig:exclusion}. We note that a recent analysis has revealed an error in earlier lattice QCD calculations of the conversion factor from $\theta$ to neutron EDM \cite{abramczyk_lattice_2017}, and updated calculations~\cite{alexandrou_neutron_2021, bhattacharya_contribution_2021} report larger uncertainties in this parameter. Our result independently excludes ALP particles in this mass range using a system that is unaffected by the theoretical uncertainty in the neutron EDM sensitivity to $\theta$.

In summary, we have performed the first precision measurements to constrain oscillatory T violation using octupolar nuclei in a crystal, leading to a bound on the parameter space of wavelike dark matter spanning 8 orders of magnitude in axionlike particle mass. Improved experimental precision is anticipated in the future, from apparatus upgrades and lower-noise detection methods.

\begin{acknowledgments}

\textit{Acknowledgment--} We thank Joseph Thywissen, Yoshiro Takahashi, Jonathan Weinstein and Andrew Jayich for helpful discussions. Bob Amos and Paul Woitalla provided technical support. M.F. acknowledges funding from a CQIQC Postdoctoral Fellowship, and A.R. acknowledges funding from a CQIQC Undergraduate Summer Research Award. This project was enabled by support from the John Templeton Foundation (Grant No. 63119), the Alfred P. Sloan Foundation (Grant No. G-2023-21045), and NSERC (SAPIN-2021-00025).
\end{acknowledgments}

\textit{Data availability--} The data that support the findings of this article are openly available~\cite{fan_2026_18879939}.

\bibliographystyle{apsrev4-2}
\bibliography{alps}

@article{Klein1987,
  title = {Phase-variation technique for measurement of the $n=2$ Lamb shift in {${\mathrm{{H}e}}^{+}$} using separated oscillatory fields},
  author = {Klein, H. A. and Hagley, E. W. and Majumder, P. K. and Pipkin, F. M. and Poitzsch, M. E.},
  journal = {Phys. Rev. A},
  volume = {36},
  issue = {7},
  pages = {3494--3496},
  year = {1987},
  doi = {10.1103/PhysRevA.36.3494},
  url = {https://link.aps.org/doi/10.1103/PhysRevA.36.3494}
}

@article{bertone2018new,
  title={A new era in the search for dark matter},
  author={Bertone, Gianfranco and Tait, Tim MP},
  journal={Nature},
  volume={562},
  number={7725},
  pages={51--56},
  year={2018},
  doi={10.1038/s41586-018-0542-z},
}

@article{Hui2021,
   author = "Hui, Lam",
   title = "Wave Dark Matter", 
   journal= "Annu. Rev. Astron. and Astrophys.",
   year = "2021",
   volume = "59",
   pages = "247-289",
   doi = "https://doi.org/10.1146/annurev-astro-120920-010024",
   url = "https://www.annualreviews.org/content/journals/10.1146/annurev-astro-120920-010024",
   publisher = "Annual Reviews",
}

@Article{Collaboration2020b,
  author  = {{Planck Collaboration: N. Aghanim} and others},
  journal = {Astronomy \& Astrophysics},
  title   = {Planck 2018 results},
  year    = {2020},
  volume  = {641},
  refid   = {10.105100046361201833910},
  url     = {https://doi.org/10.1051/0004-6361/201833910},
}

@article{ramachandran_nuclear_2023,
	title = {Nuclear {T}-violation search using octopole-deformed nuclei in a crystal},
	volume = {108},
	journal = {Phys. Rev. A},
	author = {Ramachandran, H. D. and Vutha, A. C.},
	year = {2023},
	pages = {012819},
        doi = {10.1103/PhysRevA.108.012819},
    url = {https://link.aps.org/doi/10.1103/PhysRevA.108.012819},
}

@article{konz_temperature_2003,
  title = {Temperature and concentration dependence of optical dephasing, spectral-hole lifetime, and anisotropic absorption in {$\mathrm{Eu}^{3+}$}:{${\mathrm{Y}}_{2} \mathrm{SiO}_5$}},
  author = {K\"onz, Flurin and Sun, Y. and Thiel, C. W. and Cone, R. L. and Equall, R. W. and Hutcheson, R. L. and Macfarlane, R. M.},
  journal = {Phys. Rev. B},
  volume = {68},
  issue = {8},
  pages = {085109},
  numpages = {9},
  year = {2003},
  month = {Aug},
  doi = {10.1103/PhysRevB.68.085109},
  url = {https://link.aps.org/doi/10.1103/PhysRevB.68.085109}
}

@article{cruzeiro_characterization_2018,
  title = {Characterization of the hyperfine interaction of the excited $^{5}\mathrm{D}_{0}$ state of {${\mathrm{Eu}}^{3+}:{\mathrm{Y}}_{2}{\mathrm{SiO}}_{5}$}},
  author = {Cruzeiro, Emmanuel Zambrini and Etesse, Jean and Tiranov, Alexey and Bourdel, Pierre-Antoine and Fr\"owis, Florian and Goldner, Philippe and Gisin, Nicolas and Afzelius, Mikael},
  journal = {Phys. Rev. B},
  volume = {97},
  issue = {9},
  pages = {094416},
  numpages = {13},
  year = {2018},
  month = {Mar},
  doi = {10.1103/PhysRevB.97.094416},
  url = {https://link.aps.org/doi/10.1103/PhysRevB.97.094416}
}

@Article{zhang_precision_2020,
  author  = {Zhang,S. and Lučić,N. and Galland,N. and Le Targat,R. and Goldner,P. and Fang,B. and Seidelin,S. and Le Coq,Y.},
  journal = {Appl. Phys. Lett.},
  title   = {Precision measurements of electric-field-induced frequency displacements of an ultranarrow optical transition in ions in a solid},
  year    = {2020},
  number  = {22},
  pages   = {221102},
  volume  = {117},
  doi     = {10.1063/5.0025356},
  url     = {https://pubs.aip.org/aip/apl/article/117/22/221102/39172/Precision-measurements-of-electric-field-induced},
}

@incollection{macfarlane_coherent_1987,
  title={Coherent transient and holeburning spectroscopy of rare earth ions in solids},
  author={Macfarlane, R. M. and Shelby, R. M.},
  booktitle={Spectroscopy of Solids Containing Rare Earth Ions},
  editor={Kaplyanskii, A. A. and Macfarlane, R. M.},
  volume={21},
  pages={51--184},
  year={1987},
  url={https://doi.org/10.1016/B978-0-444-87051-3.50009-2},
}

@article{foster_revealing_2018,
	title = {Revealing the dark matter halo with axion direct detection},
	volume = {97},
	issn = {2470-0010, 2470-0029},
	url = {https://link.aps.org/doi/10.1103/PhysRevD.97.123006},
	doi = {10.1103/PhysRevD.97.123006},
	number = {12},
	journal = {Physical Review D},
	author = {Foster, Joshua W. and Rodd, Nicholas L. and Safdi, Benjamin R.},
	month = jun,
	year = {2018},
	pages = {123006},
}

@article{abel_search_2017,
	title = {Search for {Axionlike} {Dark} {Matter} through {Nuclear} {Spin} {Precession} in {Electric} and {Magnetic} {Fields}},
	volume = {7},
	url = {https://link.aps.org/doi/10.1103/PhysRevX.7.041034},
	doi = {10.1103/PhysRevX.7.041034},
	number = {4},
	journal = {Physical Review X},
	author = {Abel, C. and others},
	year = {2017},
	pages = {041034},
}

@article{vanderplas_understanding_2018,
	title = {Understanding the {Lomb}–{Scargle} {Periodogram}},
	volume = {236},
	issn = {0067-0049, 1538-4365},
	url = {https://iopscience.iop.org/article/10.3847/1538-4365/aab766},
	doi = {10.3847/1538-4365/aab766},
	number = {1},
	journal = {The Astrophysical Journal Supplement Series},
	author = {VanderPlas, Jacob T.},
	month = may,
	year = {2018},
	pages = {16},
}

@article{roussy_experimental_2021,
	title = {Experimental {Constraint} on {Axionlike} {Particles} over {Seven} {Orders} of {Magnitude} in {Mass}},
	volume = {126},
	issn = {0031-9007, 1079-7114},
	url = {https://link.aps.org/doi/10.1103/PhysRevLett.126.171301},
	doi = {10.1103/PhysRevLett.126.171301},
	number = {17},
	journal = {Physical Review Letters},
	author = {Roussy, Tanya S. and Palken, Daniel A. and Cairncross, William B. and Brubaker, Benjamin M. and Gresh, Daniel N. and Grau, Matt and Cossel, Kevin C. and Ng, Kia Boon and Shagam, Yuval and Zhou, Yan and Flambaum, Victor V. and Lehnert, Konrad W. and Ye, Jun and Cornell, Eric A.},
	month = apr,
	year = {2021},
	pages = {171301},
}

@article{yano_nonlinear_1992,
	title = {Nonlinear laser spectroscopy of {Eu}$^{3+}$:{Y}$_2${SiO}$_5$ and its application to time-domain optical memory},
	volume = {9},
	issn = {0740-3224, 1520-8540},
	shorttitle = {Nonlinear laser spectroscopy of {Eu}{\textasciicircum}3+},
	url = {https://opg.optica.org/abstract.cfm?URI=josab-9-6-992},
	doi = {10.1364/JOSAB.9.000992},
	number = {6},
	journal = {Journal of the Optical Society of America B},
	author = {Yano, Ryuzi and Mitsunaga, Masaharu and Uesugi, Naoshi},
	month = jun,
	year = {1992},
	pages = {992},
}

@article{centers_stochastic_2021,
	title = {Stochastic fluctuations of bosonic dark matter},
	volume = {12},
	issn = {2041-1723},
	url = {https://www.nature.com/articles/s41467-021-27632-7},
	doi = {10.1038/s41467-021-27632-7},
	number = {1},
	journal = {Nature Communications},
	author = {Centers, Gary P. and Blanchard, John W. and Conrad, Jan and Figueroa, Nataniel L. and Garcon, Antoine and Gramolin, Alexander V. and Kimball, Derek F. Jackson and Lawson, Matthew and Pelssers, Bart and Smiga, Joseph A. and Sushkov, Alexander O. and Wickenbrock, Arne and Budker, Dmitry and Derevianko, Andrei},
	month = dec,
	year = {2021},
	pages = {7321},
}

@article{sushkov_schiff_2024,
	title = {Schiff moments of deformed nuclei},
	volume = {110},
	url = {https://link.aps.org/doi/10.1103/PhysRevC.110.015501},
	doi = {10.1103/PhysRevC.110.015501},
	number = {1},
	journal = {Physical Review C},
	author = {Sushkov, O. P.},
	year = {2024},
	pages = {015501},
}

@article{stadnik_axion_2014,
	title = {Axion-induced effects in atoms, molecules, and nuclei},
	volume = {89},
	url = {https://link.aps.org/doi/10.1103/PhysRevD.89.043522},
	doi = {10.1103/PhysRevD.89.043522},
	number = {4},
	journal = {Phys. Rev. D},
	author = {Stadnik, Y.V. and Flambaum, V.V},
	year = {2014},
	pages = {043522},
}

@article{zhang_search_2023,
	title = {Search for {Ultralight} {Dark} {Matter} with {Spectroscopy} of {Radio}-{Frequency} {Atomic} {Transitions}},
	volume = {130},
	issn = {0031-9007, 1079-7114},
	url = {https://link.aps.org/doi/10.1103/PhysRevLett.130.251002},
	doi = {10.1103/PhysRevLett.130.251002},
	number = {25},
	journal = {Physical Review Letters},
	author = {Zhang, Xue and Banerjee, Abhishek and Leyser, Mahapan and Perez, Gilad and Schiller, Stephan and Budker, Dmitry and Antypas, Dionysios},
	month = jun,
	year = {2023},
	pages = {251002},
}

@article{schulthess_new_2022,
	title = {New {Limit} on {Axionlike} {Dark} {Matter} {Using} {Cold} {Neutrons}},
	volume = {129},
	issn = {0031-9007, 1079-7114},
	url = {https://link.aps.org/doi/10.1103/PhysRevLett.129.191801},
	doi = {10.1103/PhysRevLett.129.191801},
	number = {19},
	journal = {Physical Review Letters},
	author = {Schulthess, Ivo and Chanel, Estelle and Fratangelo, Anastasio and Gottstein, Alexander and Gsponer, Andreas and Hodge, Zachary and Pistillo, Ciro and Ries, Dieter and Soldner, Torsten and Thorne, Jacob and Piegsa, Florian M.},
	month = nov,
	year = {2022},
	pages = {191801},
}

@article{flambaum_enhanced_2020,
	title = {Enhanced nuclear {Schiff} moment in stable and metastable nuclei},
	volume = {101},
	url = {https://link.aps.org/doi/10.1103/PhysRevC.101.015502},
	doi = {10.1103/PhysRevC.101.015502},
	number = {1},
	journal = {Phys. Rev. C},
	author = {Flambaum, V. V. and Feldmeier, H.},
	month = jan,
	year = {2020},
	pages = {015502},
}

@article{chadha-day_axion_2022,
	title = {Axion dark matter: {What} is it and why now?},
	volume = {8},
	issn = {2375-2548},
	shorttitle = {Axion dark matter},
	url = {https://www.science.org/doi/10.1126/sciadv.abj3618},
	doi = {10.1126/sciadv.abj3618},
	number = {8},
	journal = {Science Advances},
	author = {Chadha-Day, Francesca and Ellis, John and Marsh, David J. E.},
	month = feb,
	year = {2022},
	pages = {eabj3618},
}

@article{ferreira_ultra-light_2021,
	title = {Ultra-light dark matter},
	volume = {29},
	issn = {0935-4956, 1432-0754},
	url = {https://link.springer.com/10.1007/s00159-021-00135-6},
	doi = {10.1007/s00159-021-00135-6},
	number = {1},
	journal = {The Astronomy and Astrophysics Review},
	author = {Ferreira, Elisa G. M.},
	month = dec,
	year = {2021},
	pages = {7},
}

@misc{ohare_cajohareaxionlimits_2020,
	title = {cajohare/{AxionLimits}: {AxionLimits}},
	copyright = {Open Access},
	shorttitle = {cajohare/{AxionLimits}},
	publisher = {Zenodo},
	author = {O'Hare, Ciaran},
	month = jul,
	year = {2020},
	doi = {10.5281/ZENODO.3932429},
}

@misc{Supplemental,
  note = {See Supplemental Material for additional analysis details, which includes Refs. \cite{longdell_hyperfine_2002,yano_nonlinear_1992,stone_table_2014, derevianko_detecting_2018, baxter_recommended_2021}.}
}

@article{sherrill_analysis_2023,
	title = {Analysis of atomic-clock data to constrain variations of fundamental constants},
	volume = {25},
	issn = {1367-2630},
	url = {https://iopscience.iop.org/article/10.1088/1367-2630/aceff6},
	number = {9},
	journal = {New Journal of Physics},
	author = {Sherrill, Nathaniel and Parsons, Adam O and Baynham, Charles F A and Bowden, William and Anne Curtis, E and Hendricks, Richard and Hill, Ian R and Hobson, Richard and Margolis, Helen S and Robertson, Billy I and Schioppo, Marco and Szymaniec, Krzysztof and Tofful, Alexandra and Tunesi, Jacob and Godun, Rachel M and Calmet, Xavier},
	month = sep,
	year = {2023},
	pages = {093012},
        doi = {10.1088/1367-2630/aceff6},
}

@article{kobayashi_search_2022,
	title = {Search for {Ultralight} {Dark} {Matter} from {Long}-{Term} {Frequency} {Comparisons} of {Optical} and {Microwave} {Atomic} {Clocks}},
	volume = {129},
	issn = {0031-9007, 1079-7114},
	url = {https://link.aps.org/doi/10.1103/PhysRevLett.129.241301},
	number = {24},
	journal = {Physical Review Letters},
	author = {Kobayashi, Takumi and Takamizawa, Akifumi and Akamatsu, Daisuke and Kawasaki, Akio and Nishiyama, Akiko and Hosaka, Kazumoto and Hisai, Yusuke and Wada, Masato and Inaba, Hajime and Tanabe, Takehiko and Yasuda, Masami},
	month = dec,
	year = {2022},
	pages = {241301},
        doi = {10.1103/PhysRevLett.129.241301},
}

@article{kennedy_precision_2020,
	title = {Precision {Metrology} {Meets} {Cosmology}: {Improved} {Constraints} on {Ultralight} {Dark} {Matter} from {Atom}-{Cavity} {Frequency} {Comparisons}},
	volume = {125},
	issn = {0031-9007, 1079-7114},
	shorttitle = {Precision {Metrology} {Meets} {Cosmology}},
	url = {https://link.aps.org/doi/10.1103/PhysRevLett.125.201302},
	number = {20},
	journal = {Physical Review Letters},
	author = {Kennedy, Colin J. and Oelker, Eric and Robinson, John M. and Bothwell, Tobias and Kedar, Dhruv and Milner, William R. and Marti, G. Edward and Derevianko, Andrei and Ye, Jun},
	month = nov,
	year = {2020},
	pages = {201302},
        doi = {10.1103/PhysRevLett.125.201302},
}

@article{hees_searching_2016,
	title = {Searching for an {Oscillating} {Massive} {Scalar} {Field} as a {Dark} {Matter} {Candidate} {Using} {Atomic} {Hyperfine} {Frequency} {Comparisons}},
	volume = {117},
	copyright = {http://link.aps.org/licenses/aps-default-license},
	issn = {0031-9007, 1079-7114},
	url = {https://link.aps.org/doi/10.1103/PhysRevLett.117.061301},
	number = {6},
	journal = {Physical Review Letters},
	author = {Hees, A. and Guéna, J. and Abgrall, M. and Bize, S. and Wolf, P.},
	month = aug,
	year = {2016},
	pages = {061301},
        doi = {10.1103/PhysRevLett.117.061301},
}

@article{banerjee_oscillating_2023,
  title = {Oscillating Nuclear Charge Radii as Sensors for Ultralight Dark Matter},
  author = {Banerjee, Abhishek and Budker, Dmitry and Filzinger, Melina and Huntemann, Nils and Paz, Gil and Perez, Gilad and Porsev, Sergey and Safronova, Marianna},
  journal = {Physical Review Letters},
  volume = {135},
  issue = {22},
  pages = {223001},
  numpages = {7},
  year = {2025},
  doi = {10.1103/37vw-gc1r},
  url = {https://link.aps.org/doi/10.1103/37vw-gc1r}
}

@article{longdell_hyperfine_2002,
	title = {Hyperfine interaction in ground and excited states of praseodymium-doped yttrium orthosilicate},
	volume = {66},
	copyright = {http://link.aps.org/licenses/aps-default-license},
	issn = {0163-1829, 1095-3795},
	url = {https://link.aps.org/doi/10.1103/PhysRevB.66.035101},
	doi = {10.1103/PhysRevB.66.035101},
	number = {3},
	journal = {Physical Review B},
	author = {Longdell, J. J. and Sellars, M. J. and Manson, N. B.},
	month = jun,
	year = {2002},
	pages = {035101},
}

@techreport{stone_table_2014,
	address = {Vienna},
	title = {Table of {Nuclear} {Magnetic} {Dipole} and {Electric} {Quadrupole} {Moments}},
	month = feb,
	year = {2014},
	url = {https://www-nds.iaea.org/publications/indc/indc-nds-0658.pdf},
	number = {INDC(NDS)--0658},
	institution = {International Atomic Energy Agency (IAEA)},
	author = {Stone, N.J.},
}

@article{kim_probing_2024,
	title = {Probing an ultralight {QCD} axion with electromagnetic quadratic interaction},
	volume = {109},
	issn = {2470-0010, 2470-0029},
	url = {https://link.aps.org/doi/10.1103/PhysRevD.109.015030},
	doi = {10.1103/PhysRevD.109.015030},
	number = {1},
	journal = {Physical Review D},
	author = {Kim, Hyungjin and Lenoci, Alessandro and Perez, Gilad and Ratzinger, Wolfram},
	month = jan,
	year = {2024},
	pages = {015030},
}

@article{Blum:2014vsa,
    author = "Blum, Kfir and D'Agnolo, Raffaele Tito and Lisanti, Mariangela and Safdi, Benjamin R.",
    title = "{Constraining Axion Dark Matter with Big Bang Nucleosynthesis}",
    doi = "10.1016/j.physletb.2014.07.059",
    journal = "Phys. Lett. B",
    volume = "737",
    pages = "30--33",
    year = "2014"
}

@article{Balkin:2022qer,
    author = "Balkin, Reuven and Serra, Javi and Springmann, Konstantin and Stelzl, Stefan and Weiler, Andreas",
    title = "{White dwarfs as a probe of exceptionally light QCD axions}",
    reportNumber = "IFT-UAM/CSIC-22-136",
    doi = "10.1103/PhysRevD.109.095032",
    journal = "Phys. Rev. D",
    volume = "109",
    number = "9",
    pages = "095032",
    year = "2024"
}

@article{Hook:2017psm,
    author = "Hook, Anson and Huang, Junwu",
    title = "{Probing axions with neutron star inspirals and other stellar processes}",
    doi = "10.1007/JHEP06(2018)036",
    journal = "JHEP",
    volume = "06",
    pages = "036",
    year = "2018"
}

@article{Dine:1982ah,
    author = "Dine, Michael and Fischler, Willy",
    editor = "Srednicki, M. A.",
    title = "{The Not So Harmless Axion}",
    reportNumber = "UPR-0201T",
    doi = "10.1016/0370-2693(83)90639-1",
    journal = "Phys. Lett. B",
    volume = "120",
    pages = "137--141",
    year = "1983"
}

@article{Abbott:1982af,
    author = "Abbott, L. F. and Sikivie, P.",
    editor = "Srednicki, M. A.",
    title = "{A Cosmological Bound on the Invisible Axion}",
    reportNumber = "PRINT-82-0695 (BRANDEIS)",
    doi = "10.1016/0370-2693(83)90638-X",
    journal = "Phys. Lett. B",
    volume = "120",
    pages = "133--136",
    year = "1983"
}

@article{Preskill:1982cy,
    author = "Preskill, John and Wise, Mark B. and Wilczek, Frank",
    editor = "Srednicki, M. A.",
    title = "{Cosmology of the Invisible Axion}",
    reportNumber = "HUTP-82-A048, NSF-ITP-82-103",
    doi = "10.1016/0370-2693(83)90637-8",
    journal = "Phys. Lett. B",
    volume = "120",
    pages = "127--132",
    year = "1983"
}

@article{Hook:2018jle,
    author = "Hook, Anson",
    title = "{Solving the Hierarchy Problem Discretely}",
    doi = "10.1103/PhysRevLett.120.261802",
    journal = "Phys. Rev. Lett.",
    volume = "120",
    number = "26",
    pages = "261802",
    year = "2018",
    url = "https://link.aps.org/doi/10.1103/PhysRevLett.120.261802",
}

@article{DiLuzio:2021pxd,
    author = "Di Luzio, Luca and Gavela, Belen and Quilez, Pablo and Ringwald, Andreas",
    title = "{An even lighter QCD axion}",
    doi = "10.1007/JHEP05(2021)184",
    journal = "JHEP",
    volume = "05",
    pages = "184",
    year = "2021",
    url = "https://link.springer.com/article/10.1007/JHEP05(2021)184",
}

@article{derevianko_detecting_2018,
	title = {Detecting dark-matter waves with a network of precision-measurement tools},
	volume = {97},
	issn = {2469-9926, 2469-9934},
	url = {https://link.aps.org/doi/10.1103/PhysRevA.97.042506},
	doi = {10.1103/PhysRevA.97.042506},
	number = {4},
	journal = {Physical Review A},
	author = {Derevianko, Andrei},
	month = apr,
	year = {2018},
	pages = {042506},
}

@article{baxter_recommended_2021,
	title = {Recommended conventions for reporting results from direct dark matter searches},
	volume = {81},
	issn = {1434-6044, 1434-6052},
	url = {https://link.springer.com/10.1140/epjc/s10052-021-09655-y},
	doi = {10.1140/epjc/s10052-021-09655-y},
	number = {10},
	journal = {The European Physical Journal C},
	author = {Baxter, D. and Bloch, I. M. and Bodnia, E. and Chen, X. and Conrad, J. and Di Gangi, P. and Dobson, J. E. Y. and Durnford, D. and Haselschwardt, S. J. and Kaboth, A. and Lang, R. F. and Lin, Q. and Lippincott, W. H. and Liu, J. and Manalaysay, A. and McCabe, C. and Morå, K. D. and Naim, D. and Neilson, R. and Olcina, I. and Piro, M. -C. and Selvi, M. and Von Krosigk, B. and Westerdale, S. and Yang, Y. and Zhou, N.},
	month = oct,
	year = {2021},
	pages = {907},
}

@article{tian_reconfiguration_2011,
	title = {Reconfiguration of spectral absorption features using a frequency-chirped laser pulse},
	volume = {50},
	copyright = {https://doi.org/10.1364/OA\_License\_v1\#VOR},
	issn = {0003-6935, 1539-4522},
	url = {https://opg.optica.org/abstract.cfm?URI=ao-50-36-6548},
	doi = {10.1364/AO.50.006548},
	number = {36},
	urldate = {2025-08-01},
	journal = {Applied Optics},
	author = {Tian, Mingzhen and Chang, Tiejun and Merkel, Kristian D. and Randall, W.},
	month = dec,
	year = {2011},
	pages = {6548},
}

@misc{cheng_private_communication,
  author = "Cheng, Lan",
  howpublished = "private communication."
}

@article{catena_novel_2010,
	title = {A novel determination of the local dark matter density},
	volume = {2010},
	url = {https://dx.doi.org/10.1088/1475-7516/2010/08/004},
	doi = {10.1088/1475-7516/2010/08/004},
	number = {08},
	journal = {Journal of Cosmology and Astroparticle Physics},
	author = {Catena, Riccardo and Ullio, Piero},
	month = aug,
	year = {2010},
	pages = {004},
}

@book{ramsey_molecular_1985,
	address = {Oxford : New York},
	series = {The {International} series of monographs on physics},
	title = {Molecular beams},
	isbn = {978-0-19-852021-4},
	publisher = {Clarendon Press ; Oxford University Press},
	author = {Ramsey, Norman},
	year = {1985},
}

@article{nima_precision_2025,
  title = {Precision comagnetometry for {T}-violation searches in crystals},
  author = {Nima, Bassam and Fan, Mingyu and Radak, Aleksandar and Jayich, Andrew M. and Vutha, Amar},
  journal = {Physical Review A},
  volume = {112},
  issue = {3},
  pages = {L030801},
  numpages = {6},
  year = {2025},
  month = {Sep},
  publisher = {American Physical Society},
  doi = {10.1103/jykv-rsd1},
  url = {https://link.aps.org/doi/10.1103/jykv-rsd1}
}

@article{fuchs_searching_2025,
  title = {Searching for Dark Matter with the $^{229}\mathrm{Th}$ Nuclear Lineshape from Laser Spectroscopy},
  author = {Fuchs, Elina and Kirk, Fiona and Madge, Eric and Paranjape, Chaitanya and Peik, Ekkehard and Perez, Gilad and Ratzinger, Wolfram and Tiedau, Johannes},
  journal = {Phys. Rev. X},
  volume = {15},
  issue = {2},
  pages = {021055},
  numpages = {15},
  year = {2025},
  month = {May},
  publisher = {American Physical Society},
  doi = {10.1103/PhysRevX.15.021055},
  url = {https://link.aps.org/doi/10.1103/PhysRevX.15.021055}
}

@article{abramczyk_lattice_2017,
  title = {Lattice calculation of electric dipole moments and form factors of the nucleon},
  author = {Abramczyk, M. and Aoki, S. and Blum, T. and Izubuchi, T. and Ohki, H. and Syritsyn, S.},
  journal = {Phys. Rev. D},
  volume = {96},
  issue = {1},
  pages = {014501},
  numpages = {23},
  year = {2017},
  month = {Jul},
  publisher = {American Physical Society},
  doi = {10.1103/PhysRevD.96.014501},
  url = {https://link.aps.org/doi/10.1103/PhysRevD.96.014501}
}

@article{alexandrou_neutron_2021,
  title = {Neutron electric dipole moment using lattice QCD simulations at the physical point},
  author = {Alexandrou, C. and Athenodorou, A. and Hadjiyiannakou, K. and Todaro, A.},
  journal = {Phys. Rev. D},
  volume = {103},
  issue = {5},
  pages = {054501},
  numpages = {17},
  year = {2021},
  month = {Mar},
  publisher = {American Physical Society},
  doi = {10.1103/PhysRevD.103.054501},
  url = {https://link.aps.org/doi/10.1103/PhysRevD.103.054501}
}

@article{bhattacharya_contribution_2021,
  title = {Contribution of the QCD $\mathrm{\ensuremath{\Theta}}$-term to the nucleon electric dipole moment},
  author = {Bhattacharya, Tanmoy and Cirigliano, Vincenzo and Gupta, Rajan and Mereghetti, Emanuele and Yoon, Boram},
  journal = {Phys. Rev. D},
  volume = {103},
  issue = {11},
  pages = {114507},
  numpages = {29},
  year = {2021},
  month = {Jun},
  publisher = {American Physical Society},
  doi = {10.1103/PhysRevD.103.114507},
  url = {https://link.aps.org/doi/10.1103/PhysRevD.103.114507}
}

@dataset{fan_2026_18879939,
  author       = {Fan, Mingyu and
                  Nima, Bassam and
                  Vutha, Amar},
  title        = {Wideband search for axionlike dark matter using
                   octupolar nuclei in a crystal
                  },
  month        = mar,
  year         = 2026,
  publisher    = {Zenodo},
  doi          = {10.5281/zenodo.18879939},
  url          = {https://doi.org/10.5281/zenodo.18879939},
}

\end{document}


\title{Supplemental Material: Wideband search for axionlike dark matter using octupolar nuclei in a crystal}
\author{Mingyu Fan}
\author{Bassam Nima}
\author{Aleksandar Radak}
\author{Gonzalo Alonso-\'Alvarez}
\author{Amar Vutha}
\affiliation{Department of Physics, University of Toronto, Toronto ON M5S 1A7, Canada}

\maketitle

\section{Pulse sequence}

\begin{figure}[h!]
    \centering
    \includegraphics[width=\linewidth]{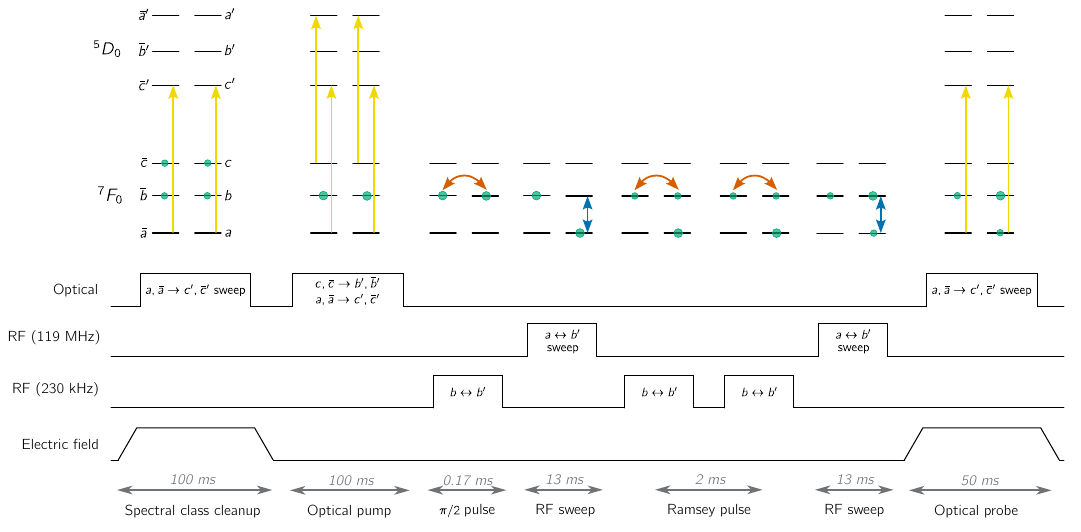}
    \caption{Experiment pulse sequence. For each step, the transitions driven are labeled as arrows, and the population distribution at the end of the pulse is shown as green circles in the energy level diagram. The corresponding pulse is shown in the timing diagram. The entire sequence takes $\SI{300}{\milli\second}$ to run, and it takes about 100 ms to process and save the data after each sequence. 6 different Ramsey pulse phase offsets are used to measure the $b -\bar{b}$ transition frequency.}
    \label{fig:pulse_sequence}
\end{figure}

In this section we explain the details of the experiment pulse sequence, shown in Fig.~\ref{fig:pulse_sequence}.

\textbf{Spectral class cleanup:} We use 2-ms-long 10-MHz-range optical sweep pulses. The pulse is repeated for 50 times to ensure a flat spectroscopy background for optical probe pulses. The electric field is ramped up 5 ms before the optical sweep pulses and turned off 2 ms after the pulses, so that the electric field during the optical pulses are constant.

\textbf{Optical pump:} The laser is switched between the two optical transitions, $c, \bar{c}\to b', \bar{b}'$ and $a, \bar{a}\to c', \bar{c}'$ to prepare the population in the $b,\bar{b}$ states. Each transition is driven for 1 ms, and we repeat driving the two transition for 50 times.

\textbf{$\pi/2$ pulse:} We use a $\pi/2$ pulse on the $b-\bar{b}$ transition to equalize the population in the two states before spectroscopy. This ensures that the state preparation does not depend on results of previous experiment cycles.

\textbf{RF sweep:} We use a sweep pulse for high-fidelity population inversion~\cite{tian_reconfiguration_2011} on the $a\leftrightarrow b$ transition. The pulse sweeps through 80 kHz of frequency range in 13 ms, which is slow enough to adiabatically transfer the population between two states. 

\textbf{Ramsey pulse:} Each $\pi/2$ pulse is 0.17 ms long and separated by 1.83 ms of free-evolution time, resulting in a center-to-center duration of $T_\mrm{Ramsey} = \SI{2}{\milli\second}$.

\textbf{Optical probe:} We scan through 25 different frequencies on each of the electric-field-shifted peaks corresponding to $\pi=\pm1$ ion ensembles. Each frequency is probed for $\SI{4}{\micro\second}$, and we repeat the frequency scan for 200 times to improve the statistical sensitivity. The measured spectrum of each peak is fit with a Gaussian function to determine the optical depth~\cite{nima_precision_2025}.

\section{Calculation of $\bra{b}\hat{I} \cdot \hat{D}\ket{b}$}

We evaluate the Eu$^{3+}$ nuclear spin unit vector along the $\hat{x}$ axis for the $\ket{b}$ state, \mbox{$\bra{b}\hat{I}\cdot\hat{D}\ket{b}$}, which is used in Eq.~6 of the main text. The hyperfine and Zeeman Hamiltonian for the nuclear spin degree of freedom in the ground $^7F_0$ electronic state in Eu:YSO is
\begin{equation}
    \mathcal{H}=\vec{I}\cdot\mathbf{Q}\cdot\vec{I}+\vec{B}\cdot\mathbf{M}\cdot\vec{I},
    \label{eq:nuclear_spin_hamiltonian}
\end{equation}
where $\vec{B}\approx350$~G$ \cdot \hat{x}$ is the magnetic field, $\mathbf{Q}$ is a tensor describing the strength of the interaction between the nuclear quadrupole moment and electric field gradients in the crystal, and $\mathbf{M}$ is the gyromagnetic tensor~\cite{longdell_hyperfine_2002}. The $\mathbf{Q}$ tensor for the ground $^7F_0$ electronic state of $^{153}$Eu:YSO is reported in Ref.~\cite{yano_nonlinear_1992}. The gyromagnetic tensor $\mathbf{M}$ has only been measured for the $^7F_0$ state of $^{151}$Eu ($I=5/2$) in YSO \cite{cruzeiro_characterization_2018}. We scale the $^{151}$Eu:YSO $\mathbf{M}$ tensor by the ratio of the nuclear magnetic dipole moments of $^{153}$Eu and $^{151}$Eu~\cite{stone_table_2014} to estimate the $\mathbf{M}$ tensor for $^{153}$Eu:YSO, finding good agreement with our experimental measurements of Zeeman shifts. The nuclear spin eigenstates are obtained from diagonalization of the above Hamiltonian. The projections of the nuclear spins along the $D_1$, $D_2$, and $b$ dielectric axes of the crystal are shown in Table~\ref{table:nuclear_spin_projections}.

\begin{table} \centering
\caption{Expectation values of the nuclear spin projections in the hyperfine eigenstates.}
\label{table:nuclear_spin_projections}
\begin{tabular}{ r|ccc } 
  & $\langle\hat{I}\cdot\hat{x}\rangle$ & $\langle\hat{I}\cdot\hat{y}\rangle$ & $\langle\hat{I}\cdot\hat{z}\rangle$ \\
 \hline
 $\lvert\bar{a}\rangle$ & -0.50 & -0.32 & 0.58 \\
 $\lvert a\rangle$ & 0.50 & 0.32 & -0.58 \\
 $\lvert\bar{b}\rangle$ & -0.25 & -0.01 & 0.31 \\
 $\lvert b\rangle$ & 0.25 & 0.01 & -0.31 \\
 $\lvert\bar{c}\rangle$ & -0.02 & 0.07 & 0.11 \\
 $\lvert c\rangle$ & 0.02 & -0.07 & -0.11 \\
 \hline
\end{tabular}
\end{table}

\section{$\delta f_d$ power spectrum}

\begin{figure}[h!]
    \centering
    \includegraphics[width=\linewidth]{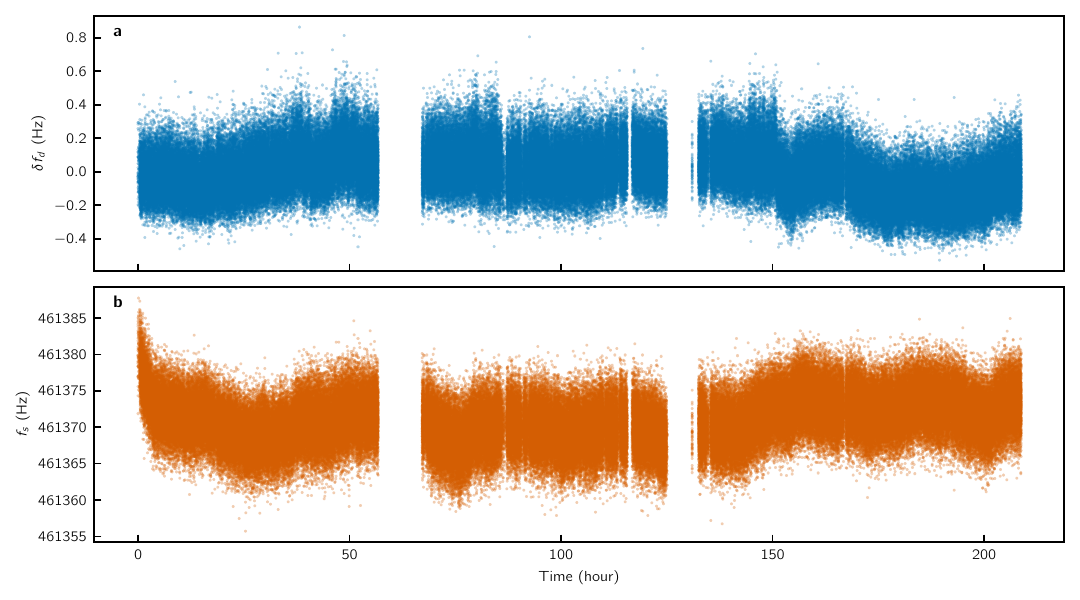}
    \caption{\textbf{a,} Time-varying frequency difference of the $b-\bar{b}$ transition between the $\pi=\pm1$ ion ensembles $\delta f_d$. This corresponds to the T-violating frequency shift from the ALP-gluon coupling. \textbf{b,} Sum frequency of the $b-\bar{b}$ transition between the $\pi=\pm1$ ion ensembles $f_s$, corresponding to the Zeeman shift. Each point in the figure represents data of one experiment cycle.
    The Zeeman shift was separated from the ALP signal using the oppositely-polarized ion ensembles as comagnetometers~\cite{nima_precision_2025}.}
    \label{fig:time_series}
\end{figure}

The time interval between $\delta f_d$ samples is irregularly spaced due to breaks in the experiment and variations of the data saving time. Some experiment breaks are visible in Fig.~\ref{fig:time_series}.

For such data with non-uniformly spaced timestamps, the Fourier transform method no longer gives the correct power spectrum. Therefore we apply the generalized Lomb-Scargle periodogram method. This method is equivalent to fitting a sinusoidal wave with a fixed frequency to the data to determine the oscillation power at the frequency~\cite{vanderplas_understanding_2018}. For uniformly spaced datapoints, the generalized Lomb-Scargle periodogram produces the same power spectrum as calculated by the Fourier transform method.

We calculate the periodogram with frequency bin size equal to the inverse of the experiment integration time. Within each bin, oscillation powers  are computed at 5 equally-spaced frequency points to ensure that each spectral peak is sufficiently sampled, and their average is used as the oscillation power of this frequency bin. The Nyquist frequency of non-uniformly spaced samples is $\sim1$ MHz, only set by the time measurement precision ($\sim\SI{1}{\micro\second}$), but our experiment bandwidth is also limited by the duration of the spectroscopy pulse ($T_\mrm{Ramsey} =\SI{2}{\milli\second}$), limiting the upper frequency bound to 500 Hz~ \cite{vanderplas_understanding_2018}. A related analysis method probing fast oscillations of a dark matter field through aliased power in a power spectrum of a low Nyquist frequency has been proposed by~\cite{derevianko_detecting_2018}.

\section{ALP signal model}

For testing ALP presences in our data, we use a model $\mathcal{M}$ to describe the $\delta f_d$ power spectrum from both an ALP signal and the background noise. The ALP signal model has two parameters, ALP mass $m_a$ and the T-violation amplitude in terms of the QCD $\theta$ parameter, $\theta_0$.

Each $\delta f_d$ is measured using six Ramsey pulses, with an average measurement frequency of $f_\mrm{cycle}=\SI{0.42}{\hertz}$. If the ALP field oscillation frequency, $f_\mrm{alp}=m_a c^2/h$, is much lower than $f_\mrm{cycle}$, the ALP-induced $\theta$ value is nearly constant during each $\delta f_d$ measurement. However, if $f_\mrm{alp}$ is comparable or higher than $f_\mrm{cycle}$, $\theta$ changes across the six Ramsey pulses, leading to undersampling of the ALP oscillations in the experimentally measured $\delta f_d$ (labeled as $\delta f_d^\mrm{exp}$). This undersampling effect reduces the $\delta f_d^\mrm{exp}$ oscillation power compared to the oscillation power of $\delta f_d$ calculated using Eqs. 3 and 6 in the main text (labeled as $\delta f_d^\mrm{theory}$). We define $\eta(f_\mrm{alp})$ as the ratio between oscillation power of $\delta f_d^\mrm{exp}$ at $f_\mrm{alp}$ and that of $\delta f_d^\mrm{theory}$, and correct the ALP signal model by multiplying the $\delta f_d^\mrm{theory}$ oscillation power by $\eta(f_\mrm{alp})$.

\begin{figure}[h!]
    \centering
    \includegraphics[width=\linewidth]{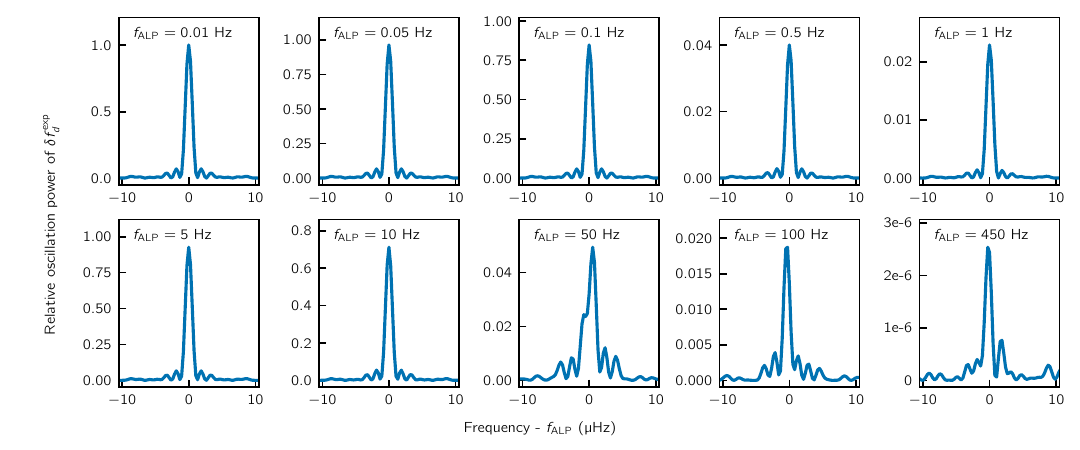}
    \caption{Calculated power spectrum of $\delta f_d^\mrm{exp}$ normalized by the peak oscillation power of $\delta f_d^\mrm{theory}$ for different ALP field oscillation frequencies $f_\mrm{ALP}$. The peak height is the correction factor of the ALP signal model, $\eta(f_\mrm{alp})$. The spectral peak width of each peak is $\SI{1}{\micro\hertz}$, in agreement of the frequency resolution of the experiment. For $f_\mrm{alp}\ll f_\mrm{cycle}$, $\delta f_d^\mrm{exp}$ peak power is unity, as the ALP field phase is nearly constant over an experiment cycle. For $f_\mrm{alp}\approx1/T_\mrm{Ramsey}$, $\delta f_d^\mrm{exp}$ peak power approaches zero as the ALP field oscillates for a full cycle during the spectroscopy pulse.}
    \label{fig:alp_model}
\end{figure}

We obtain the correction factor $\eta(f_\mrm{alp})$ using the following procedure: For each $f_\mrm{ALP}$, we calculate the spectroscopy signals $A(f, \phi, \pi=\pm1)$ that would be observed in our experiment with an ALP field. The hardware-recorded Ramsey pulse timestamps ($\SI{1}{\micro\second}$ relative precision) are used in this calculation to accurately model the experimental sensitivity to the ALP field. From the calculated spectroscopy signals, we compute $\delta f_d^\mrm{exp}$ and its power spectrum using the same analysis method as the experiment data. Then $\eta(f_\mrm{alp})$ can be determined by the oscillation power of $\delta f_d^\mrm{exp}$ and $\delta f_d^\mrm{theory}$ at $f_\mrm{alp}$. See Fig.~\ref{fig:alp_model} for example power spectra from ALP fields at different frequencies.

\begin{figure}[h!]
    \centering
    \includegraphics[width=\linewidth]{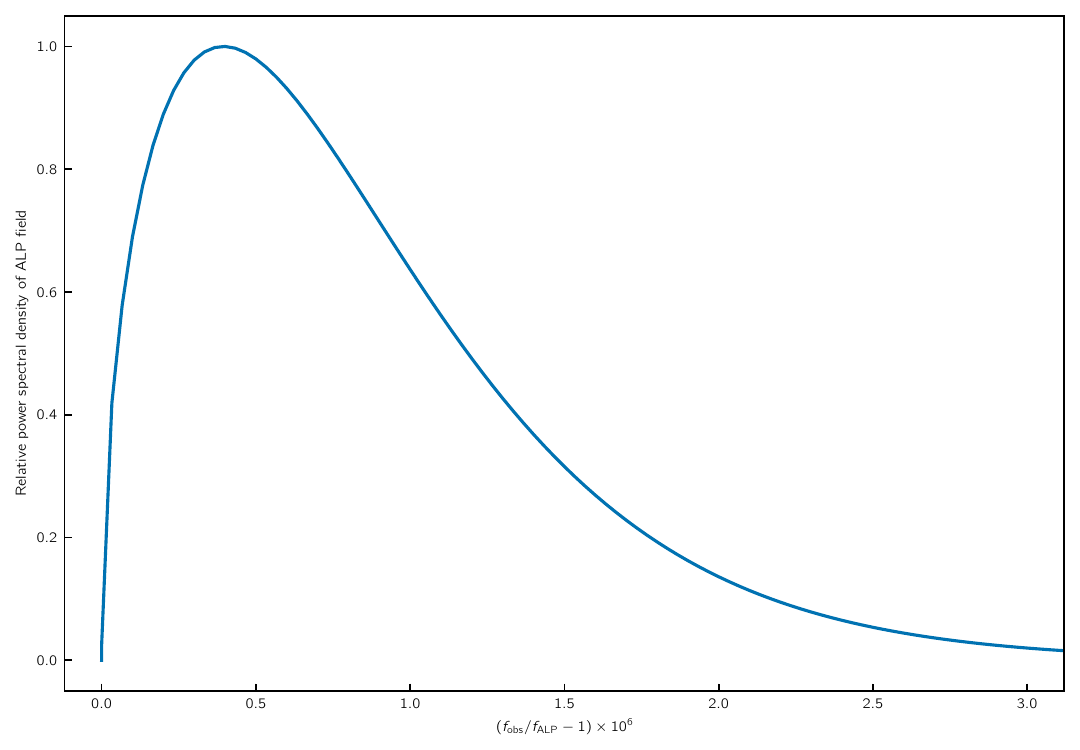}
    \caption{Power spectral density of an ALP field vs. observed ALP field oscillation frequency by the experiment, $f_\mrm{obs}$, calculated using dark matter astrophysical parameters from~\cite{baxter_recommended_2021}. The peak height is normalized to 1 for clarity.}
    \label{fig:alp_lineshape}
\end{figure}

In addition to the $\eta(f_\mrm{alp})$ dependence on $f_\mrm{ALP}$, ALP dark matter produces a distinct spectral shape that is also accounted for in this model. Using the Standard Halo Model parameters in~\cite{baxter_recommended_2021}, we model the ALP signal lineshape, shown in Fig.~\ref{fig:alp_lineshape}. For high frequency ALPs with their lineshape resolved by the experiment ($f_\mrm{alp}\gtrapprox3$ Hz), we distribute the $\delta f_d$ power of the model in frequency bins according to its lineshape.

In summary, the ALP signal model converts from $\theta$ to $\delta f_d$ using Eqs. 3 and 6 in the main text with a correction factor $\eta(f_\mrm{alp})$, and accounts for lineshape of resolved ALPs.

\section{ALP signal peak significance test}

Our likelihood analysis closely follows the procedures in~\cite{foster_revealing_2018}. To evaluate the significance of an ALP signal of a given mass, we define a test statistic
\begin{equation}
    \mrm{TS}_{\hat{\theta}_0}(m_a) = 2 [\log \mathcal{L}(d|\mathcal{M}, \{m_a, \hat{\theta}_0, \hat{\mathcal{B}}\}) - \log \mathcal{L}(d|\mathcal{M}, \{m_a, \theta_0=0, \hat{\mathcal{B}}\})].
\end{equation}

\begin{figure}[h!]
    \centering
    \includegraphics[width=\linewidth]{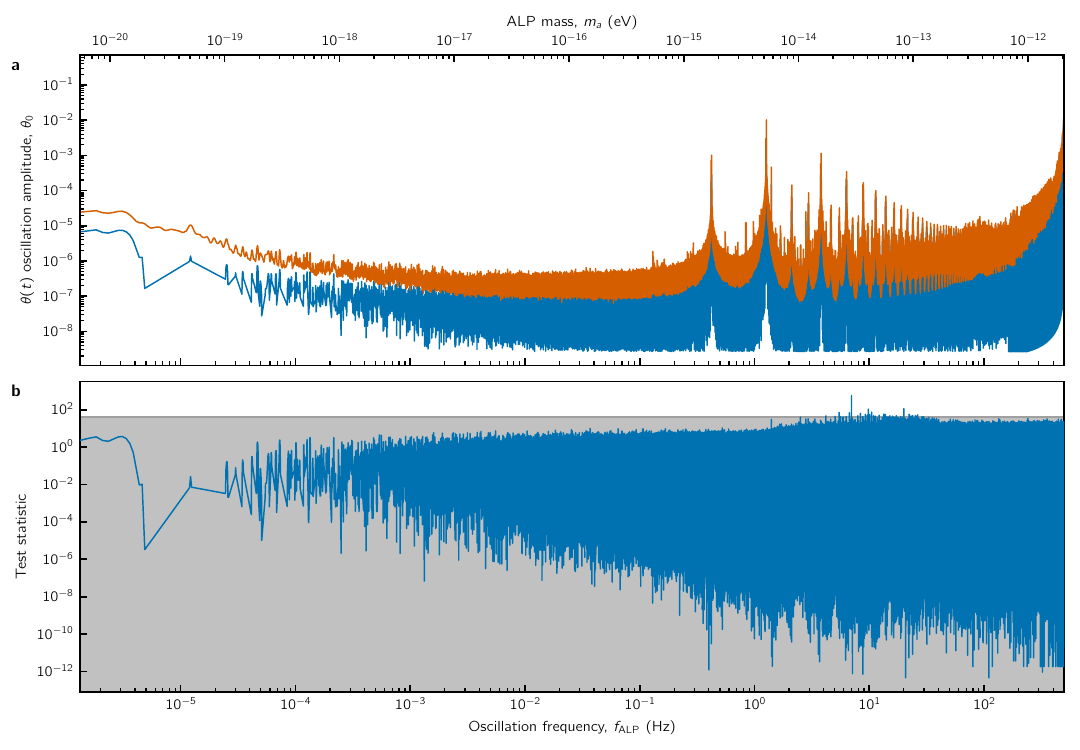}
     \caption{\textbf{a,} Maximum likelihood estimate $\hat{\theta}_0$ (blue) and 95\% upper bound of $\theta_0^{95\%}$ (orange). The increase of both quantities above 100 Hz represents reduced experiment sensitivity due to averaging of the fast-oscillating ALP field during the spectroscopy pulse. \textbf{b,} Test statistic of $\hat{\theta}_0$ (blue) and the global $3\sigma$ threshold of the test statistic. Test statistic peaks above the $3\sigma$ threshold are from technical noises, which are shown in more details in Fig.~\ref{fig:ts_zoom}.}
    \label{fig:theta}
\end{figure}

\begin{figure}[h!]
    \centering
    \includegraphics[width=\linewidth]{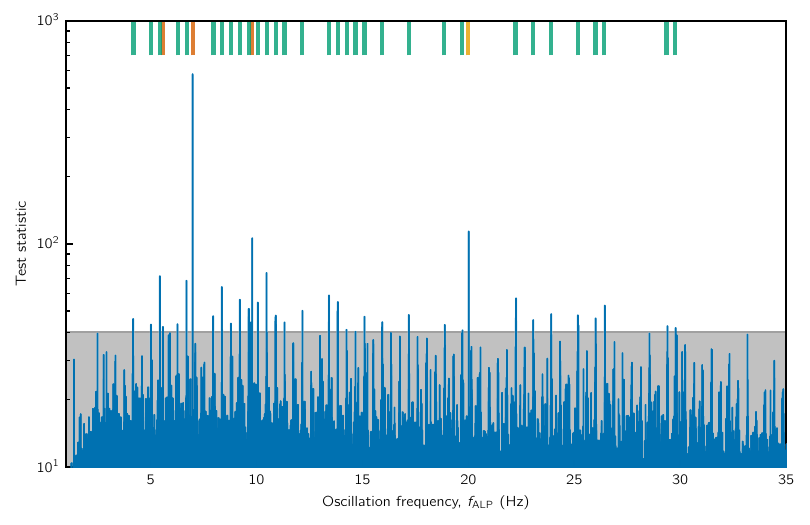}
     \caption{Test statistic of $\hat{\theta}_0$ (blue) and the global $3\sigma$ threshold (gray) between 1 Hz to 35 Hz. There are no peak above the global $3\sigma$ threshold beyond this frequency range. The technical noise frequencies are noted by vertical lines at the top: harmonics of the pulse tube cycle frequency $f_\mrm{pt}=\SI{1.40017}{\hertz}$ (orange), aliased low-frequency noise in $\delta f_d$ at multiples of the experiment cycle frequency $f_\mrm{cycle}$ (green), and data acquisition system noise at 20.0000 Hz (yellow).}
    \label{fig:ts_zoom}
\end{figure}

Higher test statistic represents higher likelihood for a signal peak to match an ALP signal, assuming that the noise model accurately describes the non-ALP background noise. To evaluate the likelihood of the ALP signal, we compare the test statistic with a global detection threshold

\begin{equation}
    \mrm{TS}_\mrm{threshold} = \left[\Phi^{-1}\left(1-\frac{p}{N_{m_a}}\right)\right]^2,
\end{equation}
where $\Phi$ is the cumulative distribution function of the standard normal distribution, $p$ is the survival function of the normal distribution ($p=1.35\times10^{-3}$ for $3\sigma$ confidence level detection), and $N_{m_a}$ is the number of distinct ALP models that we test, accounting for the look-elsewhere effect. For an experiment with resolved ALP lineshape, the number of distinct ALP models is~\cite{foster_revealing_2018}

\begin{equation}
    N_{m_a} = \frac{1}{\alpha v_0^2}\log \left(\frac{f_\mrm{max}}{f_\mrm{min}}\right),
\end{equation}
where $\alpha\approx3/4$, and $v_0$ is the speed of the local rotation curve. However, for this experiment, the ALP lineshape is not resolved below $f_\mrm{alp}\approx3$ Hz, and the number of distinct ALP models that we test in the unresolved frequency range is set by the experiment frequency resolution. Combining the resolved and unresolved frequency range, the total number of ALP models tested is $N_{m_a}=1.3\times10^7$. The test statistic and the 3$\sigma$ global detection threshold are shown in Figs.~\ref{fig:theta} and~\ref{fig:ts_zoom}. All peaks above the $3\sigma$ threshold are consistent with known technical noise frequencies of the experiment.

\section{ALP stochasticity correction factor}

If the integration time of an experiment is much shorter than the ALP field coherence time, the measured ALP field amplitude is more likely to be lower than the time-averaged value due to stochastic fluctuations of the field amplitude~\cite{centers_stochastic_2021}. As the likelihood analysis assumes that the $\theta(t)$ amplitude is constant, determined by the local density of the dark matter, the resulting bound on ALPs needs to be relaxed where the ALP coherence time is longer than the experiment duration.

\begin{figure}[h!]
    \centering
    \includegraphics[width=\linewidth]{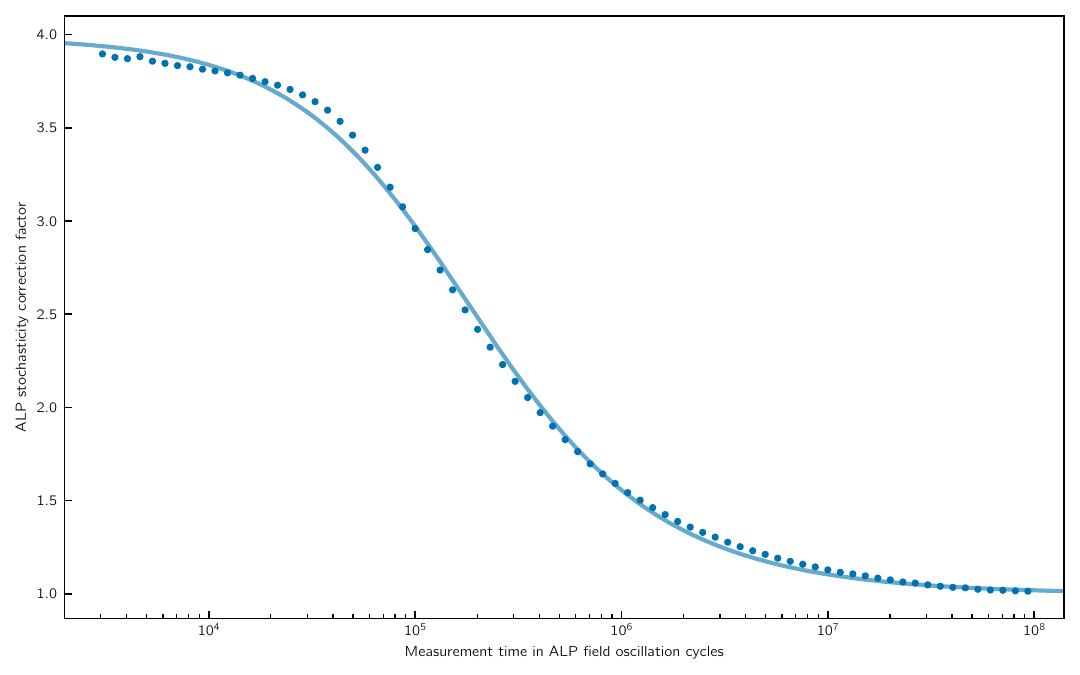}
     \caption{Stochasticity correction factors of the ALP field rms amplitude for different measurement times with a 95\% confidence level. The dots are from Monte Carlo simulations. The curve is a fit to the function $s(t) =1 + a \{\pi / 2- \tan^{-1}[(t-t_0)^b/\tau]\}$. The fitted parameters are $a=2.46$, $b=0.75$, $t_0=3.91\times10^4$, and $\tau=7.09\times10^3$. The difference between the Monte Carlo simulation results and the analytic function $s(t)$ is less than 3\%.}
    \label{fig:stochasticity}
\end{figure}

We conducted Monte Carlo simulations of the ALP field to calculate the correction factor across the ALP mass range probed by this experiment. We consider dark matter velocity up to $0.003 c$ in 500 uniformly-spaced bins. The ALP field amplitude for each velocity bin is a random variable drawn from a Rayleigh distribution with the distribution mean set by the dark matter velocity distribution using the Standard Halo Model parameters from~\cite{baxter_recommended_2021}. The ALP field phase of each bin is drawn from a uniform distribution $[0, 2\pi)$. The total ALP field is the linear sum of all ALP fields from individual velocity bins. The simulated ALP field temporal data is split into equal length of chunks, and we calculate the root-mean-squared (rms) field amplitude of each chunk. The 95\% stochasticity correction factor is calculated by dividing the rms amplitude over the entire dataset by the 95th-percentile rms amplitude computed from all chunks, sorted from largest to smallest. We repeat this procedure for different chunk lengths to compute the correction factor shown in Fig.~\ref{fig:stochasticity}.

For a short measurement time that is much shorter than the ALP coherence time, we find the constraint on the coupling strength is underestimated by a factor of 3.9 assuming that the ALP field amplitude is constant (deterministic), in agreement with the factor of $\sim3$ reported in Ref.~\cite{centers_stochastic_2021}. For measurement times that are significantly greater than the ALP coherence time, the experiment samples many different stochastically-varied ALP amplitudes and therefore the correction factor approaches 1. The experimental constraint on $C_G/f_a$ is scaled by this correction factor to account for the stochastic fluctuations of the ALP amplitude. For the convenience of other groups working on ALP experiments, we fit a simple analytic function $s(t)$ to the data, as shown in Fig.~\ref{fig:stochasticity}. This function can be directly used to correct for stochastic fluctuations in experiments that couple to the ALP field amplitude, without having to redo laborious Monte Carlo simulations each time.

\bibliographystyle{apsrev4-2}
\bibliography{alps}